# Semianalytical treatment of current density of particles injected by a monoenergetic source

P.R. Goncharov[1], B.V. Kuteev[2], V.Yu. Sergeev[1], T. Ozaki[3], S. Sudo[3]

[1] *Department of Plasma Physics, Saint Petersburg Polytechnic University, 29 Polytechnicheskaya Street, 195251, Russia*
[2] *NRC Kurchatov Institute, 1 Academician Kurchatov Square, Moscow, 123182, Russia*
[3] *National Institute for Fusion Science, 322-6 Oroshi-cho, Toki-shi, 509-5292, Japan*

**Abstract.** This paper extends the semianalytical treatment of fast ion current density to take into account time dependence and velocity diffusion using solutions of Boltzmann equation with a complete Coulomb collision term (Goncharov et al. 2010 *Phys. Plasmas* **17** 112313). An arbitrary angle distribution of the fast ion source is assumed. The results are applicable to multi-ion-species plasma. Since analytical results provide an extra physical insight and constitute a reliable basis for verification of numerical codes, it is desirable to obtain exact solutions where possible. New results clarify the discrepancies between analytical formulae in earlier bibliography and improve the physical basis of neutral beam injection current drive.

## 1. Introduction

The problem concerning the current density of fast particles originating from a monoenergetic source in a magnetized plasma arises in the context of physical basis of non-inductive electric current drive by neutral beam injection discussed in [1-3]. This problem is related to a kinetic problem concerning the distribution function of suprathermal ions and its time evolution. Non-inductive current drive is particularly important in spherical tokamaks because of the very limited space for central solenoid [4,5]. Neutral beam injection is expected to be a major non-inductive current source in ITER [6]. Numerical codes computing the beam-driven current considered in [7] as a basis for the integrated modeling for ITER, such as ONETWO [8], ACCOME [9], ASTRA [10], are using analytical solutions [11]. In NBEAMS module [12] of the ITER systems code SUPERCODE calculation of the current driven by the fast ions is also based on [11].

Formulae given in [1-3] are not applicable to multi-ion-species plasma, do not describe the time evolution owing to the use of steady state solutions, assuming a delta-like rather than arbitrary angle distribution of the fast ion source, and not taking into account velocity diffusion effects. It should be mentioned that the formulae used in [1-3] are noticeably different from fast ion distribution functions in [11,13]. Stationary solutions of Boltzmann equation used in [1-3] as well as solutions in [11,13] were obtained using various simplified expressions for Coulomb collision operator. As shown in [14], it is preferable to use the exact collision term in order to preserve its physical properties.

In this paper we extend the semi-analytical treatment using stationary and time-dependent solutions recently obtained in [14] with complete Coulomb collision operator conserving the number of particles, nullifying Maxwellian function at the equilibrium temperature, and correctly describing velocity diffusion in addition to effects taken into account in calculations of fast ion current density in [1-3].

Let $n_\alpha f_\alpha(\mathbf{v})$ be the distribution function of test particles of species $\alpha$ injected by a monoenergetic source expressed in dimensionless spherical polar coordinates as

$$S_\alpha(y,\zeta) = \frac{S_0}{2\pi v_0^3} \frac{1}{y^2} \delta(y-1) \mathcal{Z}(\zeta),$$  (1)

where $y = v/v_0$ is the dimensionless velocity magnitude, $v_0$ is the injection velocity, $\zeta = \cos\vartheta$ is the pitch angle cosine, $\delta(y-1)$ is the delta function, and $\mathcal{Z}(\zeta)$ denotes unity-normalized angular distribution of the source. Source function (1) is normalized to intensity $S_0$, *i.e.* the number of particles of species $\alpha$ injected into unit volume during unit time. Considering plasma in magnetic field, we assume the test particle distribution function to be azimuthally symmetric and expand it in a series of Legendre polynomials $P_n(\zeta)$

$$n_\alpha f_\alpha(y,\zeta) = \sum_{n=0}^{\infty} \phi_n(y) P_n(\zeta).$$  (2)

The density $j_{\alpha\parallel}$ of tangential electric current of particles of species $\alpha$ is determined by their tangential velocity, expressed via dimensionless variables as $v_\parallel = v_0 y \zeta$, averaged over the distribution function, so

$$j_{\alpha\parallel} = Z_\alpha e n_\alpha \langle v_\parallel \rangle = 2\pi Z_\alpha e v_0^4 \int_0^\infty y^3 dy \int_{-1}^{+1} \zeta n_\alpha f_\alpha(y,\zeta) d\zeta,$$  (3)

where $Z_\alpha$ is the electric charge number of test particles, and $e$ is the elementary charge. Substituting the distribution function (2) into (3), bringing the integration inside the summation sign, and using the fact that $P_0(\zeta) = 1$, and the property of Legendre polynomials stated in [15],

$$\int_{-1}^{+1} \zeta P_n(\zeta) d\zeta = 0 \text{ for } n > 1, \tag{4}$$

we conclude that the only nonzero summand in the series will be the term with $n = 1$. Since $\int_{-1}^{+1} \zeta P_1(\zeta) d\zeta = 2/3$, we finally obtain

$$j_{\alpha\parallel} = \frac{4\pi}{3} Z_\alpha e v_0^4 \int_0^\infty y^3 \phi_1(y) dy. \tag{5}$$

Note that current density (5) can be converted from Gaussian CGS units to SI as follows

$$j_{\alpha\parallel} [\text{A/m}^2] = \frac{10^3}{299792458} \times j_{\alpha\parallel} [\text{statA/cm}^2]. \tag{6}$$

In section 2 a brief comparison is given between the formulae from [1-3] and a resembling formula based on the analytical solution for the case without velocity diffusion from [11]. Next, in section 3 the exact formula for the case with no velocity diffusion is obtained using the distribution function from [14]. The formulae from [1-3] and [11], overviewed in section 2, are shown to be simpler particular cases of our exact result without velocity diffusion. In section 4 complete stationary and time-dependent solutions taking into account speed diffusion are discussed. These results based on [14] are applicable to multi-ion-species plasma and an arbitrary angular distribution of the source.

## 2. Overview of known stationary solutions

In order to represent analytical formulae for fast ion current density from [1-3] and a formula based on [11] in a unified system of notation, and to compare them with our results in the following sections, let us introduce a dimensionless constant

$$\varepsilon = \left(m_e/m_\alpha\right)^{1/3} \tag{7}$$

and two dimensional constants $v_c$ [cm/s] and $\tau_s$ [s]:

$$v_c = \varepsilon v_{T_e} \tag{8}$$

$$\tau_s = \frac{1}{4\pi e^4} \left(\frac{m_\alpha}{Z_\alpha}\right)^2 \frac{1}{n_e \Lambda} v_c^3, \tag{9}$$

where $m_e$ is the electron mass, $m_\alpha$ is the test particle mass, $v_{T_e} = \sqrt{2T_e/m_e}$ is the electron thermal velocity, $n_e$ is the electron density, and $\Lambda$ is Coulomb logarithm. We also need the effective charge

$$Z^{eff} = \frac{1}{n_e} \sum_i Z_i^2 n_i \qquad (10)$$

and another dimensionless constant

$$Z^{(b)} = \frac{m_\alpha}{n_e} \sum_i \frac{Z_i^2 n_i}{m_i}, \qquad (11)$$

where the summation is over all background plasma ion species.

Analytical expressions for fast ion current density given in [1-3] correspond to (5) with

$$\phi_1(y) = \frac{3}{4\pi v_0^3} \frac{S_0 \tilde{\tau}_s}{1+W_c^2} y^{\frac{m_i}{m_\alpha} \frac{Z^{eff}}{Z^{(b)}}} \left( \frac{1+W_c^3}{y^3+W_c^3} \right)^{1+\frac{1}{3}\frac{m_i}{m_\alpha}\frac{Z^{eff}}{Z^{(b)}}} H(1-y), \qquad (12)$$

where $H(1-y)$ is the unit step function,

$$\tilde{\tau}_s = \frac{3\sqrt{\pi}}{4} \frac{1}{4\pi e^4} m_\alpha^2 \frac{1}{n_e \Lambda} v_c^3 \qquad (13)$$

and

$$W_c = \left( \frac{3\sqrt{\pi}}{4} Z^{(b)} \frac{m_e}{m_i} \right)^{1/3} \frac{v_{T_e}}{v_0} . \qquad (14)$$

It should be noticed that (12)-(14) cannot be applied to a multi-ion-species plasma because subscript $i$ is not defined in that case. These formulae assume $Z_\alpha = 1$, which is basically the case, and a delta-like angle distribution of the fast ion source at the injection angle $\vartheta_0 = 0°$. An arbitrary distribution as in (1) is more practical due to a finite thickness and divergence of beams.

A noticeably different analytical expression can be obtained using a solution of Boltzmann equation from [11] neglecting velocity diffusion

$$\phi_1(y) = \frac{3}{2v_0^3} \frac{S_0 \hat{\tau}_s \zeta_0}{1+Y_c^3} y^{\frac{Z^{eff}}{Z^{(b)}}} \left( \frac{1+Y_c^3}{y^3+Y_c^3} \right)^{1+\frac{1}{3}\frac{Z^{eff}}{Z^{(b)}}} H(1-y), \qquad (15)$$

where $\zeta_0 = \cos\vartheta_0$ is the injection angle cosine,

$$\hat{\tau}_s = \frac{3\sqrt{\pi}}{4} \tau_s, \qquad (16)$$

$$Y_c = \left( \frac{3\sqrt{\pi}}{4} Z^{(b)} \right)^{1/3} \frac{v_c}{v_0} . \qquad (17)$$

A delta-like angle distribution of the fast ion source $\mathscr{Z}(\zeta) = \delta(\zeta - \zeta_0)$ is assumed in (15). A correcting factor $(2\pi)^{-1}$ needs to be applied before plugging (15) into (5) due to a different normalization condition adopted in [11].

Next, we will discuss an exact analytical expression for the current density without velocity diffusion effects. As a particular case of our exact result, we will also obtain a simplified formula for velocity range $v_{T_i} \ll v \ll v_{T_e}$, assumed in (12) from [1-3] and in (15) from [11].

## 3. Exact solution without velocity diffusion

Since analytical results provide an extra physical insight and constitute a reliable basis for verification of numerical codes, it is desirable to obtain exact solutions where possible. To clarify the discrepancies between the formulae for fast ion current density in various bibliographical sources, we will use the exact solution obtained in [14] for the case with no velocity diffusion

$$\phi_1(y) = \frac{3S_0 \tau_s \mathscr{Z}_1}{4\pi v_c^3} \frac{H(1-y)}{b(y)} \exp\left(-2\frac{v_0}{v_c} \int_y^1 \frac{c(\overline{y})}{b(\overline{y})} d\overline{y}\right), \tag{18}$$

where

$$b(y) = \frac{2m_\alpha}{n_e} \sum_\beta \frac{n_\beta Z_\beta^2}{m_\beta} \frac{y^2 v_0^2}{v_{T_\beta}^2} G\left(\frac{y v_0}{v_{T_\beta}}\right), \tag{19}$$

$$c(y) = \frac{v_c}{n_e} \sum_\beta \frac{n_\beta Z_\beta^2}{v_{T_\beta}} \left(\frac{1}{\sqrt{\pi}} \exp\left(-\frac{y^2 v_0^2}{v_{T_\beta}^2}\right) + \left(\frac{y v_0}{v_{T_\beta}} - \frac{v_{T_\beta}}{2 y v_0}\right) G\left(\frac{y v_0}{v_{T_\beta}}\right)\right), \tag{20}$$

the summation is over all species of the background plasma, $v_{T_\beta} = \sqrt{2T_\beta/m_\beta}$ are thermal velocities,

$G(z) = \frac{2}{\sqrt{\pi} z^2} \int_0^z x^2 e^{-x^2} dx$ is the Chandrasekhar function, and

$$\mathscr{Z}_1 = \int_{-1}^1 \mathscr{Z}(\zeta) P_1(\zeta) d\zeta. \tag{21}$$

For velocities much greater than thermal velocities of background plasma ions $v_{T_i} = \sqrt{2T_i/m_i}$ and much smaller than the thermal velocity of background plasma electrons $v_{T_e} = \sqrt{2T_e/m_e}$, i.e. for $v_{T_i} \ll v \ll v_{T_e}$, (19) and (20) reduce to the following simplified formulae

$$b(y) = Z^{(b)} + \frac{4}{3\sqrt{\pi}} \left(\frac{y v_0}{v_c}\right)^3, \tag{22}$$

$$c(y) = \frac{Z^{eff}}{2} \frac{v_c}{y v_0} + \frac{2\varepsilon}{3\sqrt{\pi}}. \tag{23}$$

The integral under the exponent in (18) can then be performed analytically to yield

$$\phi_1(y) = \frac{9 S_0 \tau_s}{16 \sqrt{\pi} v_0^3} \frac{\mathcal{Z}_1}{1 + Y_c^3} y^{\frac{Z^{eff}}{Z^{(b)}}} \left( \frac{1 + Y_c^3}{y^3 + Y_c^3} \right)^{1 + \frac{1}{3} \frac{Z^{eff}}{Z^{(b)}}} \exp\big(\Psi(y)\big) H(1-y), \qquad (24)$$

where

$$\Psi(y) = \varepsilon \left( \frac{4}{3\sqrt{\pi} Z^{(b)}} \right)^{2/3} \left( \frac{1}{6} \ln \left( \frac{(y+Y_c)^3}{y^3 + Y_c^3} \frac{1 + Y_c^3}{(1 + Y_c)^3} \right) + \frac{1}{\sqrt{3}} \arctan \left( \frac{2y - Y_c}{\sqrt{3} Y_c} \right) - \frac{1}{\sqrt{3}} \arctan \left( \frac{2 - Y_c}{\sqrt{3} Y_c} \right) \right). \quad (25)$$

As $\varepsilon$ given by (7) is a small parameter, the exponent in (24) is a rather weak dimensionless function close to unity: $\exp\big(\Psi(y)\big) \approx 1$.

## 4. Stationary and time-dependent solutions with speed diffusion

We now extend the semianalytical treatment to include the effects of velocity diffusion and the time dependence assuming a time-dependent fast particle source function

$$S_\alpha(y, \zeta, \tau) = \frac{S_0}{2\pi v_0^3} \frac{1}{y^2} \delta(y-1) \mathcal{Z}(\zeta) \big( H(\tau - \tau_0) - H(\tau - \tau_1) \big), \qquad (26)$$

where $\tau_0$ and $\tau_1$ are the turn on and turn off times respectively, $0 < \tau_0 < \tau_1$, and $\tau = t / \tau_s$ is the dimensionless time variable. For the numerical treatment instead of $\delta(y-1)$ we use a delta-like function $\mathcal{D}(y-1) = \frac{1}{\Delta\sqrt{\pi}} e^{-(y-1)^2/\Delta^2}$. Function $\phi_1(y, \tau)$ to be used in (5) to calculate the current density is determined by the equation

$$\frac{\partial \phi_1}{\partial \tau} = p(y) \frac{v_c^2}{v_0^2} \frac{\partial^2 \phi_1}{\partial y^2} + q(y) \frac{v_c}{v_0} \frac{\partial \phi_1}{\partial y} + r(y) \phi_1(y) + f(y, \tau), \qquad (27)$$

where

$$p(y) = \frac{v_c^3}{v_0^3} \frac{a(y)}{2 y^3}, \qquad (28)$$

$$q(y) = \frac{v_c^2}{v_0^2} \frac{b(y)}{y^2} - \frac{v_c^4}{v_0^4} \frac{a(y)}{2 y^4} + \frac{v_c^4}{v_0^4} \frac{1}{2 y^3} \frac{\partial a}{\partial y}, \qquad (29)$$

$$r(y) = \frac{v_c^3}{v_0^3} \frac{1}{y^2} \frac{\partial b}{\partial y} - 2 \frac{v_c^2}{v_0^2} \frac{c(y)}{y^2}, \qquad (30)$$

$$f(y, \tau) = \frac{3 S_0 \tau_s \mathcal{Z}_1}{4\pi v_0^3} \frac{1}{y^2} \frac{1}{\Delta\sqrt{\pi}} e^{-(y-1)^2/\Delta^2} \big( H(\tau - \tau_0) - H(\tau - \tau_1) \big). \qquad (31)$$

Functions $b(y)$ and $c(y)$ are defined by (19), (20), and

$$a(y) = \frac{2\varepsilon m_\alpha}{n_e} \sum_\beta \frac{n_\beta Z_\beta^2}{m_\beta} \frac{T_\beta}{T_e} \frac{y^2 v_0^2}{v_{T_\beta}^2} G\left(\frac{y v_0}{v_{T_\beta}}\right). \tag{32}$$

A detailed description of the difference scheme for (27) and the time-dependent solution procedure is given in [14], as well as the stationary solution taking into account the slowing-down, pitch angle scattering, and velocity diffusion.

The examples below correspond to injection of $10^{13}$ cm$^{-3}$s$^{-1}$ deuterons of energy 80 keV at the injection angle 45° (delta-like angle distribution of the source) into deuterium-tritium (1:1) target plasma with $n_e = 0.6 \times 10^{14}$ cm$^{-3}$, $T_e = 8$ keV, $T_i = 6$ keV. Fig. 1 shows functions $\phi_1(y)$ determining the integrand for current density calculation in (5). The dashed curve represents the exact analytical solution (18) obtained without velocity diffusion term. The dash-dotted curve illustrates the simplified formula (24). The solid curve shows the exact stationary solution from [14] taking into account velocity diffusion.

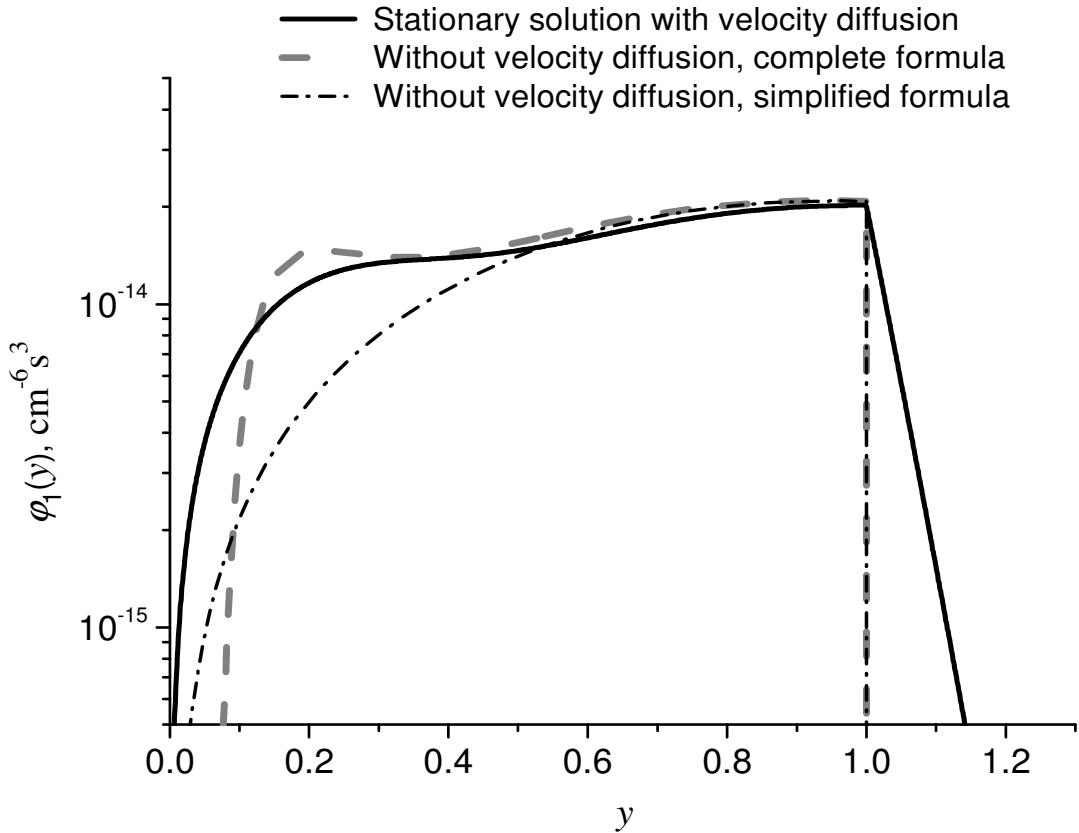

Fig. 1. Complete stationary solution taking into account velocity diffusion (solid curve), analytical solution without velocity diffusion (dashed curve), and a simplified formula (dash-dotted curve).

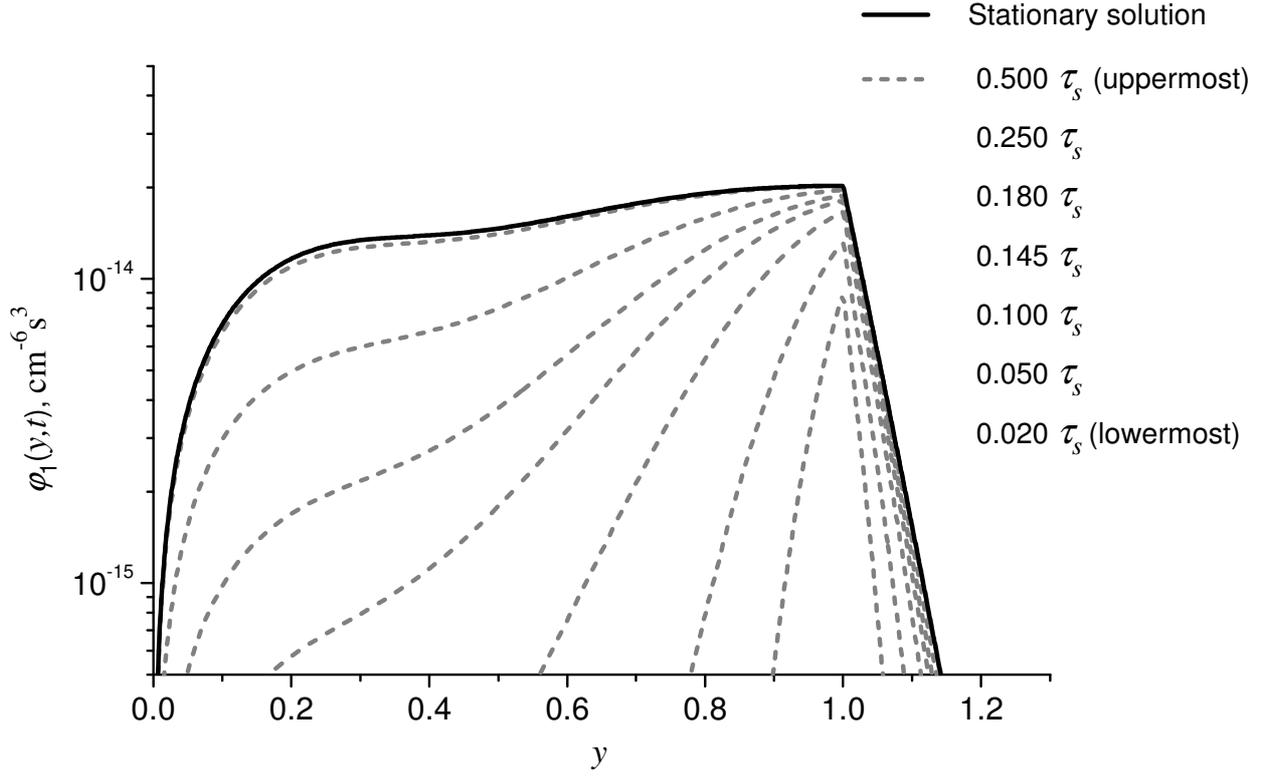

Fig. 2. Time evolution of the function determining the integrand for current density calculation.

Fig. 2 shows the solution $\phi_1(y, \tau)$ of (27) for seven time values less than $\tau_s$ (dashed curves) and the same stationary solution as in Fig. 1 (solid curve). The source turn on time was $\tau_0 = 10^{-3}$ and turn off time was $\tau_1 = 10$ in this calculation. Stationary values of the current density (5) obtained using different solutions of the kinetic equation are compared in Table 1.

Table 1. Stationary value of fast ion current density.

| Method of calculation | $j_{\alpha\parallel}$, A/m$^2$ |
|---|---|
| Analytical solution without velocity diffusion, exact formula (18) | $1.94 \times 10^5$ |
| Analytical solution without velocity diffusion, simplified formula (24) | $1.89 \times 10^5$ |
| Complete semianalytical solution with velocity diffusion | $2.19 \times 10^5$ |

## 5. Conclusion

Calculations of tangential electric current density $j_{\alpha\|}$ of test particles have been performed using recently obtained in [14] semianalytical stationary and time-dependent solutions of Boltzmann equation with Coulomb collision term and a monoenergetic source assuming an arbitrary angle distribution. Complete Coulomb collision operator is used, assuming azimuthal symmetry and Maxwellian target plasma. The solutions reflect slowing-down, velocity diffusion, and pitch angle scattering effects. Although numerical codes involving more complex models are required to obtain predictions for a particular experiment, our analysis may serve verification purposes. New results improve the physical basis of neutral beam current drive, clarify the discrepancies between analytical formulae in earlier bibliography, and extend the scope of semianalytical treatment with respect to [1-3] and [11,13].

## Acknowledgments

This work was partially supported by Rosatom Contract No. N.4b.45.03.10.1011, Contract No. 02.740.11.0468, and Project No. 2.1.1/2454 of Ministry of Education and Science of Russia. The authors are thankful to Prof. Y.N. Dnestrovskii for a discussion on the problem formulation.